\documentclass[10pt,twocolumn,aps,english,prl,superscriptaddress,showpacs]{revtex4}
\usepackage[T1]{fontenc}
\usepackage[latin9]{inputenc}
\usepackage{amsmath}
\usepackage{amssymb}
\usepackage{graphicx}
\usepackage{babel}
\usepackage{mathrsfs}

\makeatletter
\@ifundefined{textcolor}{}
{
 \definecolor{BLACK}{gray}{0}
 \definecolor{WHITE}{gray}{1}
 \definecolor{RED}{rgb}{1,0,0}
 \definecolor{GREEN}{rgb}{0,1,0}
 \definecolor{BLUE}{rgb}{0,0,1}
 \definecolor{CYAN}{cmyk}{1,0,0,0}
 \definecolor{MAGENTA}{cmyk}{0,1,0,0}
 \definecolor{YELLOW}{cmyk}{0,0,1,0}
 }
\makeatother

\begin{document}

\title{Quantum Uncertainty and Error-Disturbance Tradeoff}
\author{Yu-Xiang Zhang}
\affiliation{Hefei National Laboratory for Physical Sciences at Microscale and Department of
Modern Physics, University of Science and Technology of China, Hefei, Anhui
230026, China}
\affiliation{The CAS Center for Excellence in QIQP and the Synergetic Innovation Center
for QIQP, University of Science and Technology of China, Hefei, Anhui
230026, China}
\affiliation{Kuang Yaming Honors School, Nanjing Univeresity, Nanjing, Jiangsu 210093,
China}
\author{Shengjun Wu}
\email{sjwu@nju.edu.cn}
\affiliation{Kuang Yaming Honors School, Nanjing Univeresity, Nanjing, Jiangsu 210093,
China}
\author{Zeng-Bing Chen}
\email{zbchen@ustc.edu.cn}
\affiliation{Hefei National Laboratory for Physical Sciences at Microscale and Department of
Modern Physics, University of Science and Technology of China, Hefei, Anhui
230026, China}
\affiliation{The CAS Center for Excellence in QIQP and the Synergetic Innovation Center
for QIQP, University of Science and Technology of China, Hefei, Anhui
230026, China}
\date{\today}

\pacs{03.65.Ta, 03.65.Aa, 42.50.Xa, 03.67.-a}

\begin{abstract}
The uncertainty principle is often interpreted by the tradeoff between the
error of a measurement and the consequential disturbance to the followed
ones, which originated long ago from Heisenberg himself but now falls into
reexamination and even heated debate. Here we show that the tradeoff is
switched on or off by the quantum uncertainties of two involved
non-commuting observables: if one is more certain than the other,
there is no tradeoff; otherwise, they do have tradeoff and the Jensen-Shannon
divergence gives it a good characterization.
\end{abstract}

\maketitle


Uncertainty is an intrinsic feature of quantum mechanics that individual
particles could have no certain values with respect to a quantum
observable. A famous example is Schr\"{o}dinger's cat which stays in a
superposition of both alive and dead rather than either. Although nearly one
hundred years have passed since the dawn of quantum theory, new discovers
behind quantum uncertainty are still underway. Recent work has
illuminated quantum uncertainty's relations to quantum non-locality \cite%
{uncertainty-nonlocality}, nonclassical correlations \cite{correlation},
thermodynamics \cite{second law} and so on. For the general public, quantum
uncertainty becomes well-known because of Heisenberg's uncertainty principle
\cite{heisenberg} accompanied by the inequality $\sigma _{x}\sigma _{p}\geq
\hbar /2$ ($\sigma $ is the standard deviation). The principle is also
embodied in Robertson's inequality $\sigma _{A}\sigma _{B}\geq \frac{1}{2}%
|\langle \psi |[A,B]|\psi \rangle |^{2}$ \cite{robertson} and
Maassen-Uffink's entropic inequality $H_{\rho }(A)+H_{\rho }(B)\geq -2%
\mathrm{ln}c_{AB}$ \cite{entropy,entropy relation}. Here $c_{AB}=\mathrm{max}%
_{ij}|\langle a_{i}|b_{j}\rangle |$, where $\{|a_{i}\rangle \}$ and $%
\{|b_{j}\rangle \}$ are the eigenstates of two observables $A$ and $B$, respectively, and $%
H_{\rho }(A)$ is the Shannon entropy of the outcome distribution generated
in the ideal measurements of $A$ on ensemble $\rho $, which is determined
theoretically by Born's rule.

Following the presentation of Heisenberg himself \cite{heisenberg}, the
uncertainty principle is often interpreted in textbooks \cite{feynman} as a
consequence of the tradeoff: the higher the resolution (precision) of
measuring position is, the stronger the disturbance to particle's momentum
will be. The heart of such an interpretation positions at the back-action of
quantum measurements, as Dirac wrote \emph{\textquotedblleft a measurement
always causes the system to jump into an eigenstate of the dynamical
variable that is being measured $\cdots $\textquotedblright } \cite{dirca}.
However, these famous inequalities mentioned above do not necessarily cover
the effect of back-action and thus the relation between precision and
disturbance. To experimentally test these inequalities, one could prepare
two identical ensembles and measures the observables separately rather than
sequentially. In order to settle such a discrepancy and give a rigorous
analysis to the tradeoff relation in the mind of Heisenberg, Ozawa firstly
considered a measurement of observable $A$ occurred between the system in state $%
|\psi \rangle $ and the measurement device (the ``meter'') in state $|\xi \rangle $. As a result, he derived an
inequality \cite{ozawa} which has been verified extensively in experiments
\cite{weak measurement,neutron,other1,other2,minima,weak2}:
\begin{equation}
\epsilon _{A}\eta _{B}+\epsilon _{A}\sigma _{B}+\eta _{B}\sigma _{A}\geq
\frac{1}{2}|\langle \psi |[A,B]|\psi \rangle |.  \label{ozawa}
\end{equation}%
Here the error of this measurement is defined as $\epsilon _{A}^{2}=\langle
\psi ,\xi |(U^{\dagger }(I\otimes M)U-A\otimes I)^{2}|\psi ,\xi \rangle $,
where the meter observable is $M$, $U$ is the coupling unitary, and $I$ is
the identity operator. The consequential disturbance to the system about
observable $B$ is defined as $\eta _{B}^{2}=\langle \psi ,\xi |(U^{\dagger
}B\otimes IU-B\otimes I)^{2}|\psi ,\xi \rangle $.

Recently this relation triggered a heated debate \cite%
{proposition,qubitwenner,arxives,defi-discussion,penas,other3,buschproof,rudolph,entropic independ}%
. The controversial issue is what definition can exactly represent the
physical concepts of \emph{error} and \emph{disturbance}. The authors of
Ref. \cite{rudolph} proposed an operational criterion which requires (1) error
to be nonzero if the outcome distribution produced in an actual measurement
of $A$ deviates from that predicted by Born's rule, and (2) disturbance
to be nonzero if the back-action introduced by the actual measurement alters
the original distribution with respect to $B$. $\epsilon _{A}$ and $\eta _{B}
$ defined by Ozawa are criticized since they violate the above two requirements.
Following the requirements, the authors have showed in Ref. \cite{rudolph}
that $\langle \psi |[A,B]|\psi \rangle $ and $c_{AB}$, two characters of
uncertainty principle, are excluded to appear alone on the right hand side
of inequalities in the form of Eq. (\ref{ozawa}), leaving an open problem as
what can be there. Furthermore, Ozawa's inequality does depend on the
magnitudes of eigenvalues. But eigenvalues are not essential to
non-commuting observables and what really crucial is the family of
eigenstates. This motivates the information-theoretical approaches
(like the entropic uncertainty relation \cite{entropy,entropy relation})
that do not depend on eigenvalues. Meanwhile, some authors reported
state-independent theories \cite{qubitwenner,buschproof,entropic independ}
where the scenarios behave like benchmarking machines that just get off the
production line. The details relevant to various input states are erased in
the construction of inequalities \cite{qubitwenner,buschproof} or have never
been taken into account \cite{entropic independ}.

In spite of the intricate features, we notice the presence of $\sigma _{A}$,
$\sigma _{B}$ in Eq. (\ref{ozawa}), and the Shannon entropy $H_{\rho }(B)$
in the inequality proposed in Ref. \cite{other3}. It inspires us to consider the
relevance of \emph{quantum uncertainty} in state-dependent context. Does
quantum uncertainty play some intrinsic role behind? Consisting on
operational definitions satisfying the requirements proposed in Ref. \cite%
{rudolph}, we will show that the tradeoff between error and disturbance is
switched off or on according to the quantum uncertainties (or certainties)
of the outcome distributions of measuring $A$ and $B$. When it is switched
on, via a generally valid strategy, an inequality will be constructed to
bound their tradeoff from below.

\begin{figure}[b]
\includegraphics[width=0.46\textwidth]{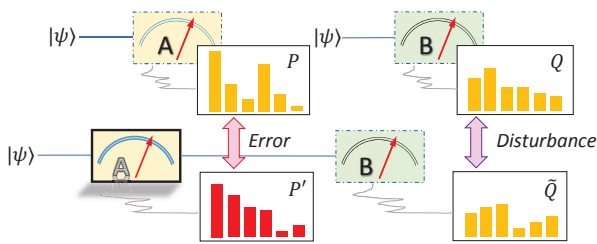}
\caption{Error and Disturbance. Dashed boxes and yellow-color bar charts
stand for the ideal measurements of which the results is determined by
Born's rule. The solid box stands for the real-life apparatus which produces
the red-color distribution $P^{\prime }$. The difference between $P$ and $%
P^{\prime }$ is caused by the imperfection of the apparatus, and the
difference between $Q$ and $\tilde{Q}$ is caused by the inevitable
back-action brought by the apparatus. Thus we define the error $Err_{\protect%
\psi }(A)$ and the disturbance $Dis_{\protect\psi }(B)$ as in Eq. (\protect
\ref{definition}).}
\label{fig.1}
\end{figure}

\noindent\textbf{Error and disturbance.}
Let us focus on the scenario illustrated in Fig. \ref{fig.1}.
By the logic of quantum mechanics,
the complete description of an isolated system is a state vector $%
|\psi \rangle $ in a Hilbert space \cite{dirca,von}, and what we obtain
from quantum measurement is a statistical distribution of different
outcomes with respect to the measured observable, as determined theoretically by Born's rule.
Suppose that Born's rule gives the distributions $P=\{p_{1},p_{2}\cdots p_{d}\}$ and $Q=\{q_{1},q_{2}\cdots q_{d}\}$, respectively, to the measurements of observables $A$ and $B$ on $|\psi \rangle $.
Then consider a real-life measurement of $A$ performed on $|\psi\rangle$,
during which the employed real-life apparatus (together with the inevitable environment)
select a preferred pointer basis in system's Hilbert space \cite{zurek}.
As a result, the entire state evolves to be $\sum_i^d\sqrt{p'_i}|a'_i\rangle_s|\phi_i\rangle_a$,
where indexes \emph{s} and \emph{a} label the system and the apparatus (possibly including the environment), and
$\langle a'_i|a'_j\rangle=\langle\phi_i|\phi_j\rangle=\delta_{ij}$.
The apparatus bridges the quantum system and the classical world we stay in.
It returns an outcome from the set ${1,2,\cdots d}$ and
finally generates the distribution $P^{\prime}=\{p_{1}^{\prime },p_{2}^{\prime }\cdots p_{d}^{\prime }\}$
with $p'_i=|\langle\psi|a'_i\rangle|^2$.
Statistically, the quantum system is then mapped to $\rho=\sum_ip'_i|a'_i\rangle\langle a'_i|$.
So the error comes from the deviation between $\{|a'_i\rangle\}_i$ and $\{|a_i\rangle\}_i$ caused by
maybe the misalignment of devices.
To quantify the error of such a real-life measurement of $A$,
we have to compare the information at hand, $P'$, to the ideal one predicted by Born's rule.
Thus, we define the error, $Err_{\psi }(A)$, to be specified below, to quantify the difference between $P$ and $P'$.

Here the modeled real-life measurements are projective measurements
that are maximally informative \cite{maximally informative}.
We do not consider the more general positive-operator valued measurements (POVM)
because fundamentally speaking, they are projective measurements performed on a larger quantum system.

The disturbance caused by the back-action of the real-life measurement of $A$
is embodied in the inequivalence between $|\psi\rangle$ and $\rho$.
With respect to observable $B$,
the disturbance displays as the difference between the distribution $Q$ and $\tilde{Q}=\{\tilde{q}_{j}=\langle b_{j}|\rho |b_{j}\rangle \}_{j=1}^{d}$,
both of which are determined by Born's rule.
If $\tilde{Q}=Q$, the induced disturbance to $B$ cannot be perceived. So we
define the disturbance term by the divergence between $\tilde{Q}$ and $Q$.

Being based on the above observation, any distance function $D(\cdot ,\cdot )$
that vanishes if and only if the two distributions are identical, such as
the relative entropy $D(P||P^{\prime })=\sum_{m}p_{m}\mathrm{ln}(p_{m}/p_{m}^{\prime })$,
can serve as a quantification of error and
disturbance. Explicitly, we quantify the error and the
disturbance $Dis_{\psi }(B)$ as
\begin{equation}
Err_{\psi }(A)=D_{1}(P,P^{\prime }),\;\;Dis_{\psi }(B)=D_{2}(Q,\tilde{Q}),
\label{definition}
\end{equation}%
where the indices \textquotedblleft 1\textquotedblright\ and
\textquotedblleft 2\textquotedblright\ mean that the two $D$-functions are
not necessary to be identical. These definitions obey the basic
operational requirements proposed in Ref. \cite{rudolph}. After the above preparation, let
us present one of our main results.

\noindent \textbf{Theorem-1.} For any $d$-dimensional pure state $|\psi
\rangle $, there exists projective measurements such that $Err_{\psi }(A)$
and $Dis_{\psi }(B)$ vanish simultaneously if and only if $P\succ Q$, i.e.,
there is no tradeoff between the error of measuring $A$ and the
consequential disturbance to $B$. $\blacksquare $

``$P\succ Q$'', read $P$ majorizes $Q$, means if sorting the elements from
larger to smaller, i.e., $p_{1}\geq p_{2}\geq \cdots p_{d}$ and $q_{1}\geq
q_{2}\geq \cdots q_{d}$, then $\sum_{i=1}^{k}p_{i}\geq \sum_{i=1}^{k}q_{i}$
for $k=1,2\cdots d$ \cite{majorization}. To give an impression, the
probability distribution of an event with certainty is in form of $(1,0,0\cdots 0)$%
, which majorizes all the other distributions; the uniform distribution $(%
\frac{1}{d},\frac{1}{d}\cdots \frac{1}{d})$ that belongs to the most
uncertain events, is majorized by all the others. By definition, $P\succ Q$
means the occurrence probability of a top-$k$ high-probability outcome in
the ideal measurement of $A$ is no less than that of $B$, for all possible
values of $k$. Thus, majorization gives a rigorous criterion for the partial
order of certainty or uncertainty. Moreover, $P\succ Q$ leads to $H(P)\leq
H(Q)$, as widely accepted that larger Shannon entropy means more
uncertainty. So Theorem-1 concludes that if the outcomes of measuring $A$ is
more certain, we are able to obtain the correct distribution of $A$ by a
projective measurement without disturbing the distribution of $B$. This
result takes a giant stride forward from the extreme case when $|\psi
\rangle $ is an eigenstate of $A$ thus $P=\{0\cdots ,1,0,\dots 0\}$ and must
majorize $Q$. We leave the technic proof of Theorem-1 in the Supplemental
Material.

It is interesting to compare Theorem-1 with Ref. \cite{rudolph} where
the authors proved the existence of $2^{d-1}$ zero-noise (zero-error) and zero-disturbance states (ZNZD) for sequential ideal measurements of $A$ and $B$.
The conclusion of Theorem-1 looks resemblant to it at the first glance.
However, the two are based on different perspectives.
Here we define error from probability distributions without considering posterior states because we cannot distinguish $|a'_i\rangle$ from $|a_i\rangle$,
with only the real-life apparatus at hand.
For fixed $A$ and $B$, Theorem-1 identifies a set of states upon which $P\succ Q$.
As a subset of system's Hilbert space, it has a non-zero measure.
Although the measure of the discrete set of the $2^{d-1}$ states is zero, the ZNZD states are defined for ideal measurements of $A$ instead of those performed with real-life apparatus.
Thus the ZNZD states can be valid for errors quantified more strictly than ours.
Additionally, maybe a coincidence, the proof of Theorem-1 supplies an algorithm to find generally also $2^{d-1}$ different realizations of $\{|a'_i\rangle\}_i$
upon which $Err_\psi(A)$ and $Dis_\psi(B)$ are null simultaneously.

\begin{figure}[bp]
\centering
\includegraphics[width=0.235\textwidth]{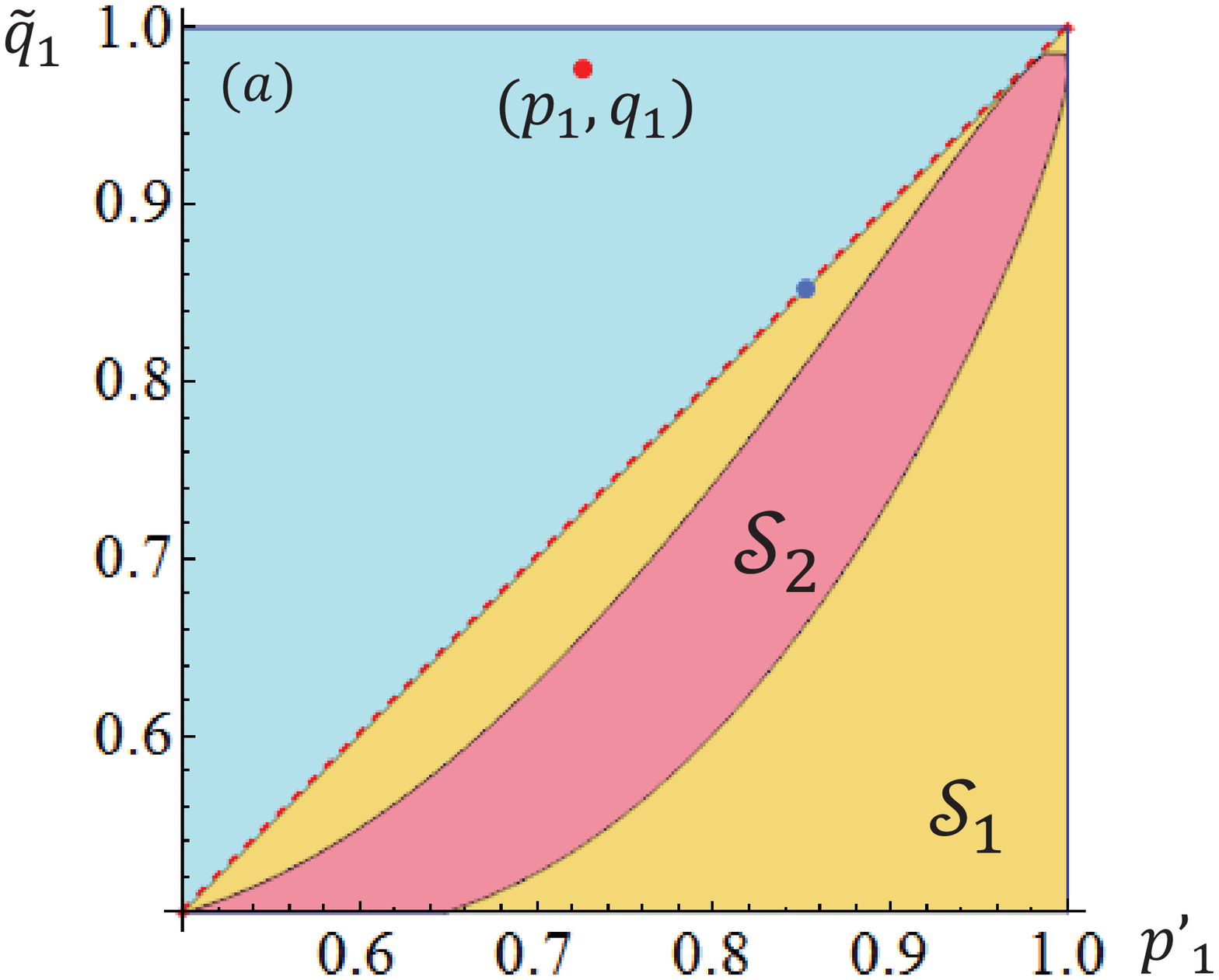}
\includegraphics[width=0.235\textwidth]{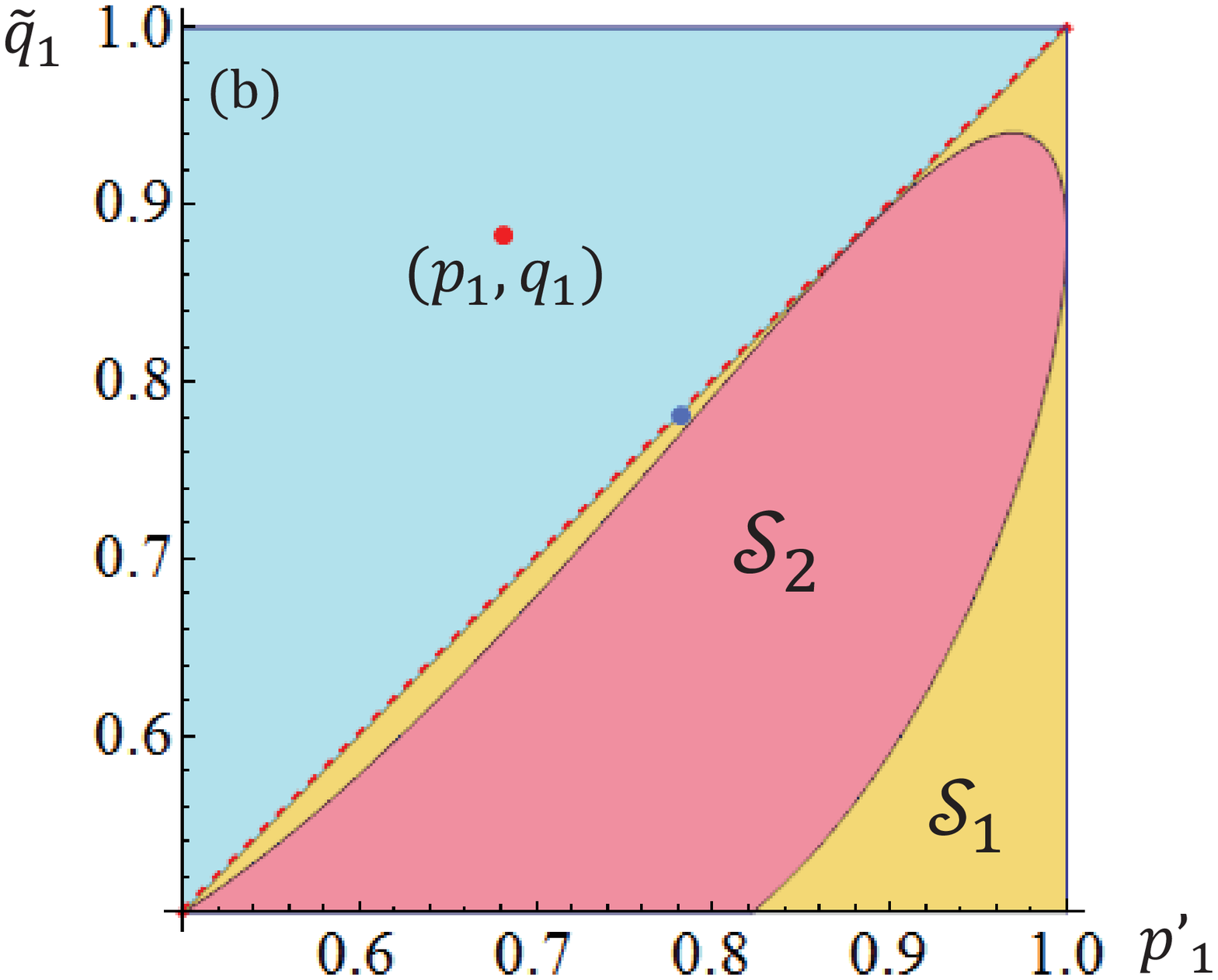}
\includegraphics[width=0.20\textwidth]{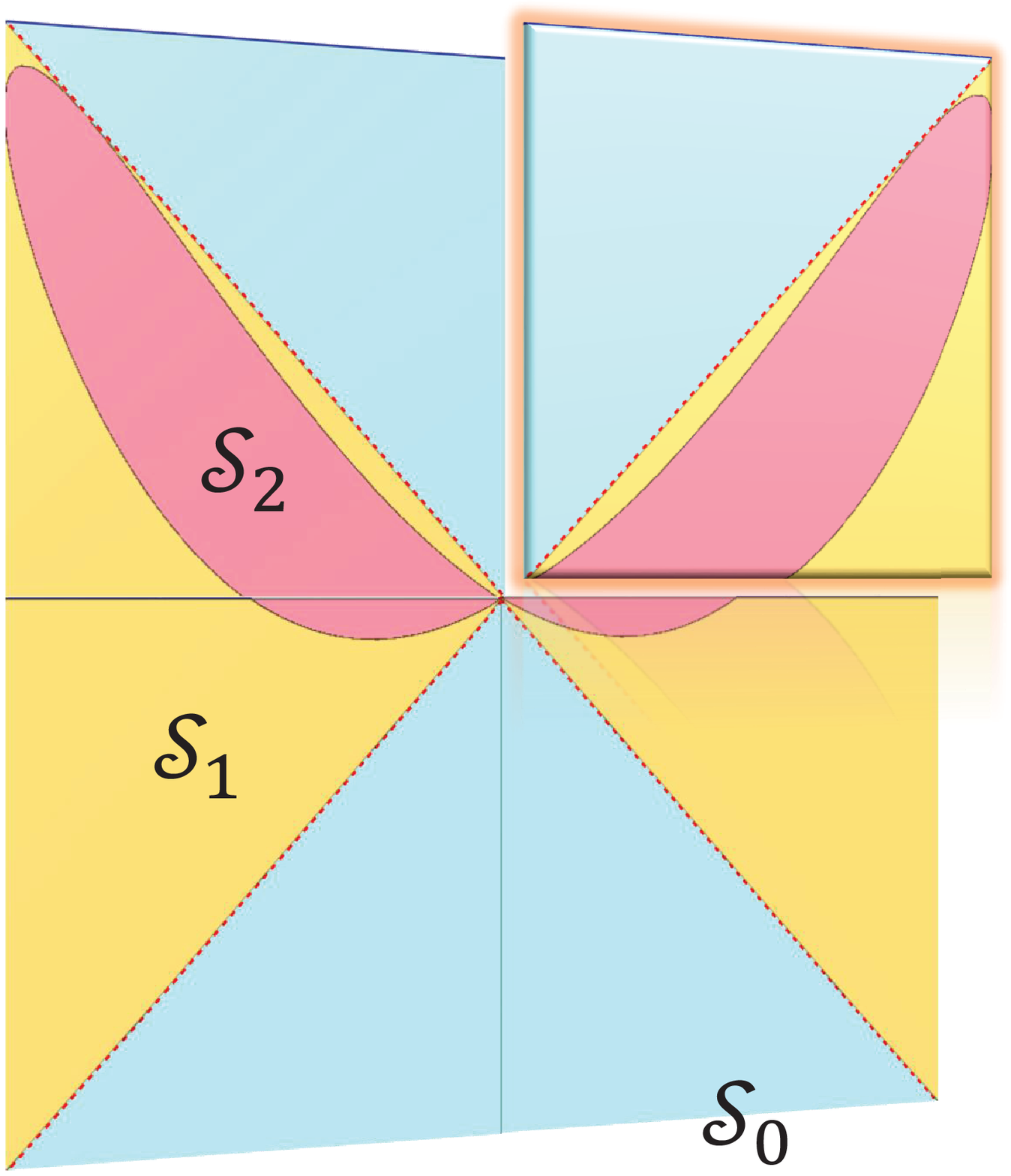}
\includegraphics[width=0.27\textwidth]{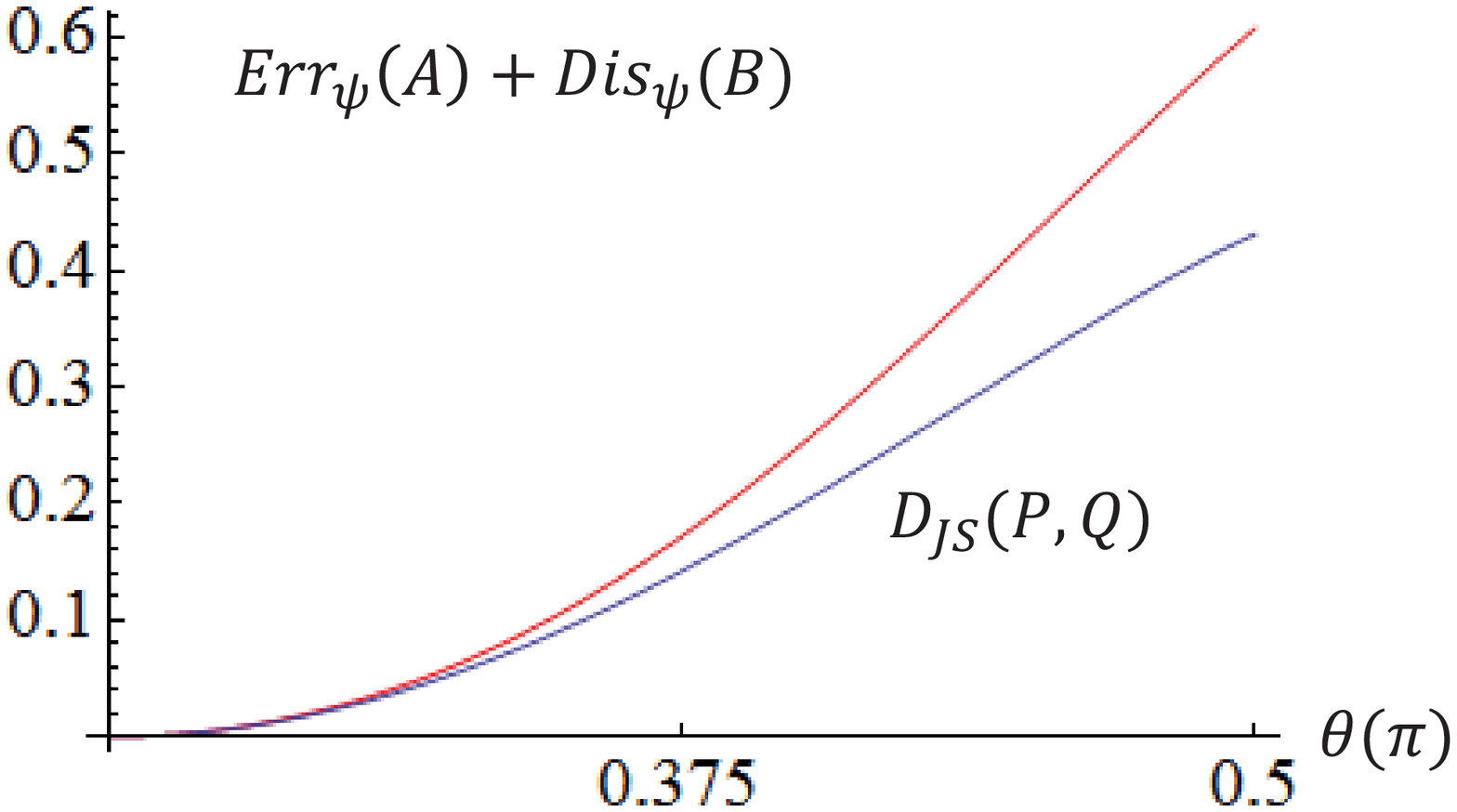}
\caption{Imbedding $\mathcal{S}_2$ and $\mathcal{S}_1$ in $\mathcal{S}_0$
when $d=2$. $\mathcal{S}_0$ is parameterized by $(p^{\prime}_1,\tilde{q}_1)$
and thus illustrated by the $[0,1]\times[0,1]$ square. We fix the labeling
of eigenstates of $A$ and $B$ such that $p_1\geq p_2$ and $q_1\geq q_2$. The
pink-shape region is $\mathcal{S}_2$, a subset of $\mathcal{S}_1$
filled with yellow. When $D$-functions in Eq. (\protect\ref{definition}) are
the relative entropy, it is sufficient to consider only the component where $%
p^{\prime}_1\geq p^{\prime}_2$ and $\tilde{q}_1\geq\tilde{q}_2$.
$(a),(b)$ show the details of that component when $P\nsucc Q$ and $(p_1,q_1)$
equals to $(0.727,0.978)$ and $(0.681,0.882)$, respectively.
The right hand side of Eq. (\protect\ref{tradeoff}) is obtained on the blue
points, which locate on the surface of $\mathcal{S}_1$.
The line-chart illustrates the exact lower bound (red) of $Err_\psi(A)+Dis_\psi(B)$ and lower bound $D_{JS}(P,Q)$ (blue) determined by Theorem-2,
when $A=\sigma_z$, $B=\sigma_x$, and $|\psi(\theta)\rangle=\cos(\theta/2)|0\rangle+\sin(\theta/2)|1\rangle$ with $\theta\in[\pi/4,\pi/2]$
such that $Q\succ P$. Here, $\sigma_z|0\rangle=|0\rangle$ and $\sigma_z|1\rangle=-|1\rangle$.}
\label{Fig.2}
\end{figure}

Next, if $P\nsucc Q$, $Err_{\psi }(A)+Dis_{\psi }(B)$ must have a positive
lower bound. To find it, we treat the sum as a functional of the probability
distribution pair $(P^{\prime },\tilde{Q})$. As coordinates, all pairs $%
(P^{\prime },\tilde{Q})=(p_{1}^{\prime },\cdots p_{d}^{\prime },\tilde{q}%
_{1}\cdots \tilde{q}_{d})$ compose a $2(d-1)$-dimensional sub-manifold $%
\mathcal{S}_{0}$ of the $2d$-dimensional Euclid space, due to the
restrictions that $\sum_{i}p_{i}^{\prime }=1$ and $\sum_{i}\tilde{q}_{i}=1$.
Meanwhile, the exact bound of $Err_{\psi }(A)+Dis_{\psi }(B)$ is the minima
over a subset of $\mathcal{S}_{0}$ defined by $p_{i}^{\prime }=|\langle
a_{i}^{\prime }|\psi \rangle |^{2}$ and $\tilde{q}_{j}=\sum_{i}p_{i}^{\prime
}|\langle a_{i}^{\prime }|b_{j}\rangle |^{2}$ (ranging over all orthogonal
basis $\{|a_{i}^{\prime }\rangle \}$), which we call $\mathcal{S}_{2}.$
Looking at the problem in $\mathcal{S}_{0}$, the target is to find the point
in $\mathcal{S}_{2}$ which is the closest to $(P,Q)\notin \mathcal{S}_{2}$,
where \textquotedblleft closest\textquotedblright\ is defined by the $D$%
-functions applied in Eq. (\ref{definition}). Analytical solution of the
exact bound seems complicated to approach and must shape terrible because of
the involved geometry of $\mathcal{S}_{2}$ when embedded in $\mathcal{S}_{0}$
(see Fig. \ref{Fig.2}).

However, we can extend $D_{1}(P,P^{\prime })+D_{2}(Q,\tilde{Q})$ naturally
from $\mathcal{S}_{2}$ to the entire $\mathcal{S}_{0}$. Define the set of
pairs satisfying $P^{\prime }\succ \tilde{Q}$ as $\mathcal{S}_{1}$.
According to Horn's theorem \cite{majorization} which says two distributions
$P^{\prime }\succ \tilde{Q}$ if and only if there is a unitary $U$ such that $%
p_{i}^{\prime }=\sum_{j}|U_{ij}|^{2}\tilde{q}_{j}$, $\mathcal{S}_{2}$ must
be a subset of $\mathcal{S}_{1}$. Thus we have $\mathcal{S}_{2}\subset
\mathcal{S}_{1}\subset \mathcal{S}_{0}$ and $\mathrm{min}_{\mathcal{S}%
_{2}}(\cdots )\geq \mathrm{min}_{\mathcal{S}_{1}}(\cdots )$ by logic. Moreover,
enormous $D$-functions make the point $(P,Q)\notin \mathcal{S}_{1}$ be
the unique extreme point of $Err_{\psi }(A)+Dis_{\psi }(B)$ in $\mathcal{S}%
_{0}$ such that its minima value over $\mathcal{S}_{1}$ must be obtained at
the surface of $\mathcal{S}_{1}$. This surface consists of many faces,
especially those defined by majorization, i.e., $P^{\prime }\succ \tilde{Q}$
is defined by $d-1$ inequalities with symbol $\geq $, the point $(P^{\prime
},\tilde{Q})$ locates on the surface if some ``$\geq $'' is actually ``$=$''. The
dimensions of these faces range from $d-1$ to $2d-3$. Although not easy, the
geometry of $\mathcal{S}_{1}$ is much simpler than that of $\mathcal{S}_{2}$
and analytical solution becomes reachable. So the generally valid strategy
is to find the minima over $\mathcal{S}_{1}$.

After stating the mathematical correctness, let us discuss the physical
implications of such a strategy. Look at Fig. \ref{fig.1}, in the part of
sequential measurements, the measurement of $B$ is designed to be an ideal
one. Image that we replace it with real-life apparatus which performs projective
measurements in basis denoted by $\{|b^{\prime}_j\rangle\}_j$. So $\tilde{Q}$
should be redefined as $\sum_ip^{\prime}_i|\langle
b^{\prime}_j|a^{\prime}_i\rangle|^2$. Then starting from Horn's theorem \cite%
{majorization}, it is not difficult to see that now the minima of $%
Err_\psi(A)+Dis_\psi(B)$ is exactly the minima over $\mathcal{S}_1$ desired
by us.

In the following we will set up the inequality when the $D$-functions are
the relative entropy, a prime concept in information theory with widely
applications but generally not so easy to handle. On the way to the final
answer, it is interesting that a new concept emerges as a nature extension
of majorization.

\noindent\textbf{Majorization by sections.} If $P\nsucc Q$, we cannot
conclude that $P$ is more certain than $Q$ in the global sense. However if
only particular outcomes are concerned, things will be different. Let us relabel
the eigenstates so that $p_1\geq p_2\cdots\geq p_d$ and $q_1\geq
q_2\cdots\geq q_d$, then cut the subscript string $1\sim d$ into short
sections $(1\sim j_1),(j_1+1\sim j_2)\cdots (j_k+1\sim d)$. For each
section, say the $t$-th one, we find out the probabilities according to the
subscripts in this section and take their sum $P_{S_t}=%
\sum_{i=j_{t-1}+1}^{j_t}p_i$ and $Q_{S_t}=\sum_{i=j_{t-1}+1}^{j_t}q_i$ ($%
j_0=1$, $j_{k+1}=d$). Then we say \emph{P majorizes Q by sections} if the
relation
\begin{equation}
\frac{1}{P_{S_t}}(p_{j_{t-1}+1},\cdots p_{j_{t}})\succ\frac{1}{Q_{S_t}}%
(q_{j_{t-1}+1},\cdots q_{j_t})  \label{section}
\end{equation}
holds for all the short sections. (If some zero-valued probabilities vanish
the denominator, take the limit from infinitesimal positive factors.) Equation (%
\ref{section}) says that $P$ is relatively more certain than $Q$ in each
section. We use $P\succ_\mathcal{P} Q$ to denote this relation where the
index $\mathcal{P}$ labels this particular partition of the subscript
string. In addition, $P_{(\mathcal{P})}=\{P_{S_t}\}_t$ and $Q_{(\mathcal{P}%
)}=\{Q_{S_t}\}_t$ are two distributions coarse-grained from $P$ and $Q$ by
this partition.

Next, let us find all the coarsest partitions under which $P$ majorizes $Q$
by sections. We say a partition is coarser than another if the later can be
obtained from the former by additional cutting such as to cut $(2\sim 9)$
into $(2\sim 4)(5\sim 9)$). (\textquotedblleft Coarser\textquotedblright\
defined in such a way is a partial order, we cannot say $(1\sim 5)(5\sim 9)$
is coarser than $(1\sim 4)(4\sim 7)(7\sim 9)$.) Let us denote the set of all
the coarsest partitions upon which $P$ majorizes $Q$ by sections as $\mathscr{P}_{PQ}$.
None of $\mathscr{P}_{PQ}$ can be obtained
by extra cutting from another one in it. $\mathscr{P}_{PQ}$ is never an
empty set since $P$ will always majorize $Q$ by sections for the finest
partition where each short section consists of only one subscript. Then
according to each partition in $\mathscr{P}_{PQ}$ (say, the one labeled by $%
\mathcal{P}$), we coarsen $P$ and $Q$ to obtain distributions $P_{(\mathcal{P%
})}=\{P_{S_{1}},P_{S_{2}}\cdots \}$ and $Q_{(\mathcal{P})}=%
\{Q_{S_{1}},Q_{S_{2}}\cdots \}$ in the way given above. With these
preparations, now we can present our tradeoff relation.

\noindent\textbf{Theorem-2.} Label the outcomes such that $p_1\geq
p_2\cdots\geq p_d$ and $q_1\geq q_2\cdots\geq q_d$. Then if $P\nsucc Q$, $%
Err_\psi(A)=D(P||P^{\prime})$ and $Dis_\psi(B)=D(Q||\tilde{Q})$, there is a tradeoff relation
\begin{equation}
Err_\psi(A)+Dis_\psi(B)\geq\underset{\mathcal{P}\in\mathscr{P}_{PQ}}{\mathrm{%
min}}D_{JS}(P_{(\mathcal{P})},Q_{(\mathcal{P})}),  \label{tradeoff}
\end{equation}
where $D_{JS}(P_1,P_2)\equiv2H(\frac{P_1+P_2}{2})-H(P_1)-H(P_2)$ is the
Jensen-Shannon divergence. Moreover, $D_{JS}(P,Q)$ serves as the lower bound
if $Q\succ P$ and $Q\succ_\mathcal{P}P$ for any possible partition $\mathcal{%
P}$. $\blacksquare$

The proof is left in the Supplemental Material. Actually there is a one-to-one
correspondence between the partitions of subscript string $(1\sim d)$
and the faces defined from $P^{\prime }\succ \tilde{Q}$ on the surface of $%
\mathcal{S}_{1}$. What Theorem-2 gives us is an algorithm of searching the
lower bound on various faces, and such searching is generally necessary since $%
\mathcal{S}_{1}$ is fixed but the point $(P,Q)$ moves case by case. The
second part of Theorem-2 simply identifies a specific situation, namely, a case in
point of such a complexity. Particularly, for qubits, $\mathcal{S}_{1}$ has a
simple geometry and there is only one partition in $\mathscr{P}_{PQ}$. So
the lower bound is straightforwardly $D_{JS}(P,Q)$ (see Fig. \ref{Fig.2}).

An evident advantage of our operational definitions is that the experimental
test can be easily performed. To test Ozawa's inequality one needs three
state method or the technology of weak measurement \cite{weak
measurement,neutron,other1,other2,minima,weak2}. While for ours, one just
needs to arrange the devices according to Fig. \ref{fig.1} with only the
input state in concern. When the qubit is coded by polarizations of photons, the
experimental configuration includes only an ordinary single-photon source,
four single-photon detectors and some wave plates.

\noindent \textbf{Conclusions and discussions.} To summarize, the theory
built in this Letter bridges the theories of the Proposition and the
Opposition in debate. The relation between error, disturbance and quantum
uncertainty, three kinds of terms appeared in Eq. (\ref{ozawa}), is clearly
described by Theorem-1 and Theorem-2, which state that if defining error and disturbance
by probability distributions, the tradeoff is switched off if $P\succ Q$ and
switched on if $P\nsucc Q$. Meanwhile, our theorems compose a
state-dependent theory that is finer than the state-independent work \cite%
{arxives} where error and disturbance are also defined by probability
distributions. Moreover, Theorem-2 tells that the Jensen-Shannon divergence and the
coarse-grained distributions serve in the lower bound, thus giving an answer
to the open question asked in Ref. \cite{rudolph}.

For further generalization, one could consider input ensembles described by
mixed states due to the lack of classical information. Some recent work
makes progress in this direction, such as tighter bounds \cite{mixed} and
separating uncertainty into quantum and classical parts \cite{q-c
uncertainty}. We show in the Supplemental Material that Theorem-2 is valid for all the mixed input states,
and Theorem-1 is robust
to depolarization noise, i.e., valid for ensembles described by
$\frac{\eta }{d}I+(1-\eta )|\psi \rangle \langle \psi |$, as well as for all
qubit states, pure or mixed. We leave the more general case as an open question.
Another interesting topic is the connection between
error-disturbance theory and the multipartite quantum correlations \cite{lili}.
We hope our work could supply or inspire new ideas and methods in the
study of quantum uncertainty and quantum measurements.

We thank Zhihao Ma, Yutaka Shikano and Xuanmin Zhu for discussions. This
work was supported by the National Natural Science Foundation of China
(Grant Nos. 11275181 and 61125502), the National Fundamental Research
Program of China (Grant No. 2011CB921300), and the CAS.

\clearpage
\section{Supplemental Material}

The Supplemental Material consists of three sections, the first two sections
give the proof of the two theorems for pure states, respectively. The
extension to mixed states is in the third section.

\noindent\textbf{Proof of Theorem-1}

The proposition that error and disturbance can be zero simultaneously is
equivalent to the proposition that there is a unitary matrix which satisfies
the following two conditions simultaneously:
\begin{equation}
\begin{cases}
\sum_ip_i|U_{ij}|^2=q_j, \\
\sum_jU_{ij}\sqrt{q_j}=\sqrt{p_i}.%
\end{cases}%
\end{equation}

Here we just need to show its sufficiency. For $B$, we have the
freedom to define the phases of its eigenstates $\{|b_j\rangle\}_{j=1}^d$ so
that the state $|\psi\rangle$ can be written as
\begin{equation}
|\psi\rangle=\sum_j\sqrt{q_j}|b_j\rangle.
\end{equation}
Then if $U_{ij}=\langle a^{\prime}_i|b_j\rangle$ satisfies the above two
conditions, we have
\begin{equation}
\begin{aligned} p'_i&=\sum_{mn}U_{im}U_{in}^{*}\sqrt{q_mq_n}=p_i,\\
\tilde{q}_j&=\sum_ip'_i|\langle a'_i|b_j\rangle|^2=q_j. \end{aligned}
\end{equation}

Then we will prove the existence of such a unitary with the premise $P\succ Q$ by
mathematical induction. (Horn's theorem states that the first conditions
has solutions if and only if $P\succ Q$.) First, when $d=2$, if $(p_1,
p_2)\succ(q_1, q_2)$ (for convenience, we assume $p_1>p_2$, the case $p_1=p_2$ is
trivial), then the following unitary is the answer:
\begin{equation}
\begin{aligned} &\frac{e^{-i\phi}}{\sqrt{p_1-p_2}}\begin{pmatrix}
\sqrt{q_1-p_2} & e^{i\theta_1}\sqrt{p_1-q_1}\\ e^{i\theta_2}\sqrt{p_1-q_1} &
-e^{i(\theta_1+\theta_2)}\sqrt{q_1-p_2} \end{pmatrix}, \\
&\phi=\mathrm{arcsin}[\sqrt{p_2(p_1-q_1)}/\sqrt{q_1(p_1-p_2)}],\\
&\theta_1=\mathrm{arcsin}[\sqrt{p_1p_2}/\sqrt{q_1q_2}],\\
&\theta_2=\phi+\mathrm{arcsin}[\sqrt{p_1(q_1-p_2)}/\sqrt{q_1(p_1-p_2)}].
\end{aligned}
\end{equation}
Actually, we will get a second solution by taking $-\phi,-\theta_1$ and $%
-\theta_2$ in the above matrix. Here, we do not require the normalization
that $\sum_ip_i=1$. Then we assume the validity of the cases where the dimension
equals to $d-1$.

For $d$-dimensional cases, the first condition can be written as
\begin{equation}
U^\dagger Diag\{p_1, p_2\cdots p_d\}U=diag\{q_1, q_2\cdots q_d\},
\end{equation}
where we use $Diag$ to denote diagonal matrix and $diag\{q_1, q_2\cdots q_d\}
$ means a matrix whose diagonal elements are $q_1, q_2\cdots q_d$.

For convenience, we assume that $p_1\geq p_2\geq \cdots\geq p_d$ and $q_1$
is the largest one in $Q$. There exists a subscript $j$ such that $p_{j-1}\geq
q_1\geq p_{j}$. Then we have $(p_1, p_j)\succ(q_1, p_1+p_j-q_1)$ and $%
(p_2\cdots p_{j-1},p_1+p_j-q_1, p_{j+1}\cdots p_d)\succ(q_2\cdots q_d)$. To
the first, majorization is valid since $p_1\geq(q_1, p_1+p_j-q_1)\geq p_j$. For the second
majorization, when $2\leq k\leq j-1$, since $p_1\geq\cdots p_{j-1}\geq{q_1}$%
, we have $\sum_{i=2}^kp_i\geq (k-1)q_1\geq\sum_{i=2}^kq_i$; when $j\leq
k\leq d-1$, we have $\sum_{i=2}^kq_i=\sum_{i=1}^kq_i-q_1\leq%
\sum_{i=1}^kp_i-q_1$. Since $q_2\geq q_3\cdots\geq q_d$, the second majorization must be valid.

Then we have a unitary $U_1$ that acts only on the subspace belongs to $p_1,
p_j$, such that it changes the diagonal elements $p_1, p_j$ to $q_1,
p_1+p_j-q_1$ and maps vector $(\sqrt{q_1}, \sqrt{p_1+p_j-q_1})$. Next,
according to our induction assumption, we have another unitary $U_2$ acting
on the subspace belongs to $p_2, p_3\cdots p_d$ that changes $%
Diag\{p_2\cdots p_{j-1}, p_1+p_j-q_1\cdots p_d\}$ to $diag\{q_2\cdots q_d\}$
and maps vector $(\sqrt{q_2}\cdots\sqrt{q_d})$ to $(\sqrt{p_2}\cdots\sqrt{%
p_1+p_j-q_1}\cdots\sqrt{p_d})$. Then $U_1U_2$ is just the unitary we want
for the $d$-dimensional cases.

Since we have two solutions when $d=2$, it can be seen from the induction that
generally $2^{d-1}$ solutions can be found.

\bigskip \noindent\textbf{Proof of Theorem-2}

Consider a $2(d-1)$-dimensional compact manifold embedded in $R^{2d}$ with
coordinates $(P^{\prime}, \tilde{Q})$ with restrictions $\sum_i
P^{\prime}_i=\sum_j\tilde{Q}_j=1$ and $P^{\prime}_i\geq0, \tilde{Q}_j\geq0$.
We denote it by $\mathcal{S}_0$. It has a $2(d-1)$-dimensional sub-manifold $%
\mathcal{S}_1$ with an additional restriction that $P^{\prime}\succ\tilde{Q}$%
. Moreover, $\mathcal{S}_1$ has a subset labeled by $\mathcal{S}_2$, in which the points $%
(P^{\prime},\tilde{Q})$ are defined through an orthonormal basis $%
\{|a^{\prime}_i\rangle\}$ that $p^{\prime}_i=|\langle\psi|a^{\prime}_i%
\rangle|^2$ and $\tilde{q}_j=\sum_i p^{\prime}_i|\langle
a^{\prime}_i|b_j\rangle|^2$. So the exact lower bound of $E_\psi(A)+D_\psi(B)
$ is the minimum value of the function $D(P||P^{\prime})+D(Q||\tilde{Q})$
over the set $\mathcal{S}_2$.

Now we extend the definition of function $D(P||P^{\prime})+D(Q||\tilde{Q})$
naturally to the entire manifold $\mathcal{S}_0$. Since $\mathcal{S}_2\subset%
\mathcal{S}_1$, we have $\mathrm{min}_{\mathcal{S}_2}(\cdots)\geq\mathrm{min}%
_{\mathcal{S}_1}(\cdots)$. As the geometry of $\mathcal{S}_2$ is too complex,
we give up the exact bound and aim at a lower bound defined by the value of
\begin{equation}
\underset{(P^{\prime},\tilde{Q})\in\mathcal{S}_1}{\mathrm{min}}%
\,D(P||P^{\prime})+D(Q||\tilde{Q}).  \label{S1}
\end{equation}
The only zero point of $E_\psi(A)+D_\psi(B)$, and the only extreme point, is
$(P,Q)$, which is outside of $\mathcal{S}_1$. From the results of
mathematical analysis, the minimum defined in Eq. (\ref{S1}) will be
obtained on the surface of $\mathcal{S}_1$.

For convenience, we make use of the freedom of relabeling to assume that $%
p_1\geq p_2\cdots\geq p_d$ and so do distributions $Q$. Then $P^{\prime}\,%
\text{and}\,\tilde{Q}$, which are also labeled by such an order, will give $%
\sum_i p_i\mathrm{ln}\frac{p_i}{p^{\prime}_i}+\sum_j q_j\mathrm{ln}\frac{q_j%
}{\tilde{q}_j}$ the smallest value.

\textbf{Lemma.} If the probability distributions $P$ and $Q$ are sorted by the order that
$p_1\geq p_2\geq\cdots\geq p_d$ and $q_1\geq q_2\cdots\geq q_d$,
then among all ways of labeling the probabilities in $P^{\prime}
$ and $\tilde{Q}$, the one satisfying $p^{\prime}_1\geq p^{\prime}_2\geq
\cdots\geq p^{\prime}_d$ and $\tilde{q}_1\geq\tilde{q}_2\cdots\geq\tilde{q}_d
$ gives the minima to $D(P||P')+D(Q||\tilde{Q})$.

Proof. We just need to prove it for $D(P||P^{\prime})$. Let us use
$\{p'^{T}_i\}$ to denote permutations other than the decreasing order.
We will show that $D(P||P^{\prime})$ is smaller than $D(P||P'^T)$.
Expanding their subtraction we have
\begin{equation}
\begin{aligned} &\sum
p_i\mathrm{ln}\frac{p_i}{p'_i}-p_i\mathrm{ln}\frac{p_i}{p'^T_i}=\sum
p_i\mathrm{ln}\frac{p'^T_i}{p'_i}\\ =&p_n\sum_{i=1}^n
\mathrm{ln}\frac{p'^T_i}{p'_i}+(p_{n-1}-p_n)\sum_{i=1}^{n-1}\mathrm{ln}%
\frac{p'^T_i}{p'_i}+\cdots\\
&+(p_t-p_{t+1})\sum_{i=1}^t\mathrm{ln}\frac{p'^T_i}{p'_i}+\cdots+(p_1-p_2)%
\mathrm{ln}\frac{p'^T_1}{p'_1} \end{aligned}
\end{equation}
Then it is direct to see that none is positive in the above expansion. $%
\blacksquare$

Without loss of generality, we assume that elements in $P$ and $Q$ are
all positive. For possible problems caused by zero elements, we can take
the limit from infinitesimal positive factors.

Consider the geometrical surface of $\mathcal{S}_1$ in manifold $\mathcal{S}%
_0$. First, it is composed by $(d!)^2$ symmetric components due to
permutation. The above lemma tells us that we just need to consider the
single component on which $P^{\prime}$ and $\tilde{Q}$ are labeled in the
decreasing order. Such a component looks like a polytope with many faces and
we just need to take the faces associated with the definition of
majorization into account. (Faces associated with some equations like $%
p^{\prime}_j=0$ or $\tilde{q}_j=0$ points on them will give infinite value
to the sum of error and disturbance, thus we do not need to care about them;
for other faces of the component associates with the decreasing order of
labeling, actually they are not the faces of $\mathcal{S}_1$.) Consider the
following equations
\begin{equation}
\sum_{i=1}^kp^{\prime}_i=\sum_{i=1}^k\tilde{q}_i\qquad (k=1,2\cdots d-1).
\label{equations}
\end{equation}
Now we use $n$ to denote the dimension of the manifold, i.e., $n=2(d-1)$. An $%
(n-j)$-dimension surfaces of $\mathcal{S}_1$ is produced by $j$ equations of
the above equation string, accompanied with the restriction that $%
P^{\prime}\succ\tilde{Q}$.

Now let us consider the minimum value on an $(n-k)$-dimension surface ($j$
ranges from 1 to $d-1$). The $j$ equations cut the subscript string $1\sim d$
into $k+1$ sections that the sum of $p^{\prime}_i$ and the sum of $\tilde{q}%
_i$ within each subscript-section are equal. We use $S_t$ to denote the
different sections and define notations
\begin{equation}
P^{\prime}_{S_t}\equiv\sum_{i\in S_t}p^{\prime}_i
\end{equation}
and so do the distributions $P, Q\, \text{and}\,\tilde{Q}$. Then we use the
Lagrange multiplier method to search for the extreme value:
\begin{equation}
\begin{aligned}
\mathcal{L}=-&\sum_i(p_i\mathrm{ln}p'_i+q_i\mathrm{ln}\tilde{q}_i)-H(P)-H(Q)%
\\
+&\sum_{t=1}^{k+1}\lambda_t(P'_{S_t}-\tilde{Q}_{S_t})+\lambda_p(%
\sum_ip'_i-1), \end{aligned}
\end{equation}
where the equivalence in each section has already implies that $\sum_i\tilde{%
q}_i=1$. Simple calculation shows that the minima is obtained on the point
\begin{equation}
\begin{aligned} p'_i=\frac{p_i}{2}(1+\frac{Q_{S_t}}{P_{S_t}})\;\;
\tilde{q}_i=\frac{q_i}{2}(1+\frac{P_{S_t}}{Q_{S_t}}) \;\; (i\in S_t)
\end{aligned}  \label{extreme point}
\end{equation}
if the subscript ``$i$'' is in the $t$-th section when $\lambda_p=2$ and $%
\lambda_t=-2Q_{S_t}/(P_{S_t}+Q_{S_t})$. To write down the minimum value, we
define two distributions obtained from $P, Q$ by coarse graining:
\begin{equation}
\begin{aligned} P_{(\mathcal{P})}=\{P_{S_1}, P_{S_2}\cdots P_{S_{k+1}}\}\\
Q_{(\mathcal{P})}=\{Q_{S_1}, Q_{S_2}\cdots Q_{S_{k+1}}\} \end{aligned}
\end{equation}
and an average of the two, $\frac{1}{2}(P_{(\mathcal{P})}+Q_{(\mathcal{P})})$
with elements $\{\frac{1}{2}(P_{S_t}+Q_{S_t})\}_{t=1}^{k+1}$. Then the
minimum value obtained from Eq. (\ref{extreme point}) is given by the
Jensen-Shannon divergence, $D_{JS}(P_{(\mathcal{P})},Q_{(\mathcal{P})})$.

Now we have to check that whether this point is located on the $(n-j)$%
-surface of $\mathcal{S}_1$, i.e., whether $(P^{\prime},\tilde{Q})$ given by
Eq. (\ref{extreme point}) satisfies the requirement that $P^{\prime}\succ%
\tilde{Q}$. Since $P^{\prime}_{S_t}=\tilde{Q}_{S_t}$, $P^{\prime}\succ\tilde{%
Q}$ if and only if $P$ and $Q$ satisfy the condition that within any
section, such as $S_t$. After re-normalizing $P_{S_t}$ and $Q_{S_t}$ to a
common factor we must have $P\succ Q$ in each section. More rigorously,
suppose that the section $S_t$ has subscripts $j_{t-1}+1,\cdots j_t$, then $%
P^{\prime}\succ\tilde{Q}$ if and only if
\begin{equation}
\frac{1}{P_{S_t}}(p_{j_{t-1}+1},\cdots p_{j_t})\succ \frac{1}{Q_{S_t}}%
(q_{j_{t-1}+1},\cdots q_{j_t})
\end{equation}
for all these sections. This is just the conception ``majorization by
sections'' introduced in the main text.

If the above condition is not satisfied, the point defined by Eq. (\ref%
{extreme point}) locates outside of $\mathcal{S}_1$ and we should consider
the edges of the $(n-j)$-dimensional surface, i.e., we should add another
equation in Eq. (\ref{equations}) and study the $(n-j-1)$-dimensional case. If the
above condition is satisfied, finer partition, i.e., adding extra equations
in Eq. (\ref{equations}), will not bring lower value.

So we have to find the family of all the coarsest partitions of the
subscript string (anyone in this family is not a refinement of another one
in it) under which $P$ majorizes $Q$ by sections. Then we calculate the
Jensen-Shannon divergence corresponding to the partitions and the minimum
one is just the minimum over $\mathcal{S}_1$. With the
notations $\mathscr{P}_{PQ}$, the above analysis leads to Theorem-2 in
the main text.

One may wonder whether the solution given by Eq. (\ref{extreme point})
follows the order $p^{\prime}_1\geq p^{\prime}_2\geq\cdots\geq p^{\prime}_d$
and $\tilde{q}_1\geq\tilde{q}_2\geq\cdots\geq\tilde{q}_d$. Actually, we do
not need to care. This is because $P\succ_\mathcal{P} Q$ ensures $%
P^{\prime}\succ\tilde{Q}$ such that the solution is in $\mathcal{S}_1$.
Thus all the values derived from $\mathscr{P}_{PQ}$ can be reached in $\mathcal{S}%
_1$, and meanwhile the real minima over $\mathcal{S}_1$ must link with one
partition in $\mathscr{P}_{PQ}$. So the minimizing over $\mathscr{P}_{PQ}$
will always give the minima we want.

The second part of Theorem-2 can be checked straightforwardly.

\bigskip \noindent\textbf{Extension to Mixed States}

Firstly, for qubit mixed states, we give a visualizable proof for
Theorem-1 with the help of Bloch-sphere. Note that the density matrix of the
input state, $A$, $B$ and the observable of the real-life measurement $O_A$%
, can be represented by four vectors $\vec{r}, \vec{a}$, $\vec{b}$ and $%
\vec{a}^{\prime}$. Then the probability distributions have one to one
correspondence with the inner products such as $\vec{r}\cdot\vec{a}$ and $%
\vec{r}\cdot\vec{b}$. They can be assumed to be positive due to the freedom of
relabeling the eigenstates of $A$ and $B$. Suppose the angle between $\vec{r}
$ and $\vec{a}$ is $\theta_a$ and the angle between $\vec{r}$ and $\vec{b}$
is $\theta_b$. Now $P\succ Q$ implies that $\theta_a\leq\theta_b$. $\vec{a}%
^{\prime}$ can be obtained by rotating $\vec{a}$ around $\vec{r}$ such that $%
P^{\prime}=P$. Thus the angle $\xi$ between $\vec{a}^{\prime}$ and $\vec{r}$
will range from $\theta_b-\theta_a$ to $\theta_a+\theta_b$. Then there
must be a case where we have $\cos{\theta_a}\cos{\xi}=\cos{\theta_b}$,
which then leads to $\tilde{Q}=Q$.

For higher dimensions, we can also represent $|\psi\rangle$, the projectors
of eigenstates of observables $A$ and $B$ by coherent vectors and generators
of the Lie-algebra of $SU(d)$ as
\begin{equation}
|\psi\rangle\langle\psi|=\frac{1}{d}(I+\sqrt{\frac{d(d-1)}{2}}\vec{r}\cdot%
\vec{\lambda}),  \label{coherent vector}
\end{equation}
where $\{\lambda_i\}_{i=1}^{d^2-1}$ are generators of the Lie-algebra. They
satisfy the restriction that $\mathrm{tr}(\lambda_i\lambda_j)=2\delta_{ij}$.
Then one can see that the norm of $\vec{r}$ is not relevant. What
does matter is the direction of $\vec{r}$ and the inter-angles between the
vectors representing $|\psi\rangle$, $|a^{\prime}_i\rangle$ and $|b_j\rangle$.
So the conclusion of our Theorem-1 in the main text is still valid for
mixed states in the form of $\frac{\eta}{d}I+\eta|\psi\rangle\langle\psi|$ whose
coherent vector is parallel with (but shorter than) $\vec{r}$, the coherent
vector of $|\psi\rangle\langle\psi|$.
When $d=2$, all the mixed states can be written in this form.
That means that Theorem-1 is valid for all qubit states, pure or mixed.

As to Theorem-2, firstly let us redefine $\mathcal{S}_2$ according to the mixed states in consideration, i.e., $P'=\{p'_i=\langle a'_i|\rho|a'_i\rangle\}$ and $\tilde{Q}=\{\tilde{q}_j=\sum_i p'_i|\langle b_j|a'_i\rangle|^2\}$.
Suppose the spectrum representation of the density matrix is $%
\rho=\sum_i\rho_i|\psi_i\rangle\langle\psi_i|$.
The eigenvalues compose a probability distribution and we denote it as $%
\varrho=(\rho_1,\rho_2\cdots\rho_d)$.
Now we can define $\tilde{\mathcal{S}}_1$,
a $2(d-1)$-dimensional sub-manifold of both $\mathcal{S}_0$ and $\mathcal{S}%
_1$, with points whose coordinates $(P^{\prime},\tilde{Q})$ satisfy the
condition such that $\varrho\succ P^{\prime}\succ\tilde{Q}$.
Since we have $%
\mathcal{S}_2\subset\tilde{\mathcal{S}}_1\subset\mathcal{S}_1$, the lower
bound given by our Theorem-2 is still valid.
Actually one can do more analysis in $\tilde{\mathcal{S}}_1$ to get a tighter bound.

\end{document}